\documentclass{aa}  
\usepackage{graphicx,natbib,ulem}
\usepackage{txfonts}
\usepackage{color}

\begin{document}
  \titlerunning{ SS 433:  Accretion revealed in H$\alpha$}
\authorrunning{ Bowler}
   \title{SS 433: The accretion disk revealed in H$\alpha$}

   \subtitle{}

   \author{M. G.\ Bowler \
          }

   \offprints{M. G. Bowler \\   \email{m.bowler1@physics.ox.ac.uk}}
   \institute{University of Oxford, Department of Physics, Keble Road,
              Oxford, OX1 3RH, UK}
   \date{Received 14 December 2009; accepted 17 April 2010}

 
  \abstract
   {The Galactic microquasar SS\,433 is very luminous and ejects opposite jets at approximately one quarter the speed of light. It is regarded as a super-Eddington accretor but until recently there were no observations of accretion. } 
  {To present an analysis of spectroscopic optical data obtained before and during  a major flare, which yield in H$\alpha$ unambiguous evidence for the accretion disk.  }
   {Already published high resolution spectra, taken with a 3.6-m telescope almost
     nightly over 0.4 of a precession cycle, are analysed. }  
  { The spectra, taken almost nightly in August and September 2004,
    revealed a period of quiescence followed by activity
    which culminated in the
    accretion disk of SS 433 becoming visible. The visible material in the
    accretion disk orbited the compact object at greater than 500 km
    s$^{-1}$, implying that the mass of the compact object is less than 37 $M_\odot$. Evidence that an accretion stream joins the disk at over 700 km s$^{-1}$ suggests that the mass is considerably below this upper limit. The accretion disk clearly orbits the centre of
    mass of the binary system with the compact object, sharing its speed of approximately 175 km s$^{-1}$. The mass of the companion lies between 20 and 30 $M_\odot$ and it probably does not fill its Roche lobe. }
{ } 

   \keywords{Stars: binaries: eclipsing -- Accretion disk }

   \maketitle
%

\section{Introduction}

The Galactic microquasar  SS 433 is famous for its continual ejection
of plasma in two opposite jets at approximately one quarter the speed
of light. The system is a binary with a period of 13.08 days
(Crampton, Cowley and Hutchings 1980) and eclipses at both conjunctions
(e.g. Goranskii et al 1998). It seems taken for granted that the compact object is a super-Eddington accretor (see Fabrika 2004 for a comprehensive review), yet until very recently there has been no direct evidence for an accretion disk surrounding the compact object. There is evidence that He II 4686 \AA\  emission has been
observed from the base of the jets (Crampton and Hutchings 1981,
Fabrika and Bychkova 1990, Fabrika 1997) and C II lines orbiting with the compact object have been detected (Gies et al 2002). The orbital speed of the compact object about the binary centre of mass is thus known to be approximately 175 km s$^{-1}$. This means the system is massive, certainly greater than 10 $M_\odot$ (see Fabrika 2004); the observations of the circumbinary disk (Blundell, Bowler \&
Schmidtobreick 2008, Perez M \& Blundell 2009) require approximately 40 $M_\odot$. 

In August 2004 nightly observations of SS 433, with the 3.6-m telescope on La Silla, commenced and continued until it became a daylight object late in
September. Results obtained on the relativistic jets  (Blundell, Bowler,
Schmidtobreick 2007) and from the
stationary emission lines during the first half of the series of
observations (Blundell, Bowler \& Schmidtobreick 2008) have already
been published. The second paper revealed the circumbinary disk surrounding the SS 433 system and alluded to unveiling of the accretion disk after JD 2453000 + 274. No analysis of those data has been published, but the H$\alpha$ spectra themselves can be found in Schmidtobreick \& Blundell (2006) Fig.2. The H$\alpha$ spectra there shown cover JD +245 to + 310 and it is evident that after JD +280 the H$\alpha$ structure broadens and in particular a prominent feature appears to the blue, first emerging clearly on JD +288. This deviates ever bluewards for about half a period and I contend that it constitutes a dramatic signature of the accretion disk of SS 433. In the sections below I discuss the published spectra and my analysis in terms of Gaussian components. I then present the evidence that these spectra revealed the accretion disk of SS 433 and the inferences drawn from these quantitative data.

\section{Data}

 During the period to JD +274 SS 433 was
quiescent; thereafter  increased activity culminated in a flaring episode around JD + 294 (Blundell, Bowler \& Schmidtobreick
2007, 2008). Between +274.5 and +287.5 there were only two observations
(+281.5, +282.5) and there was a gap of two nights after
291.5. During the quiescent period the H$\alpha$ line could for the
 most part be represented by three components only, namely a broad component
 identified with the wind from the accretion disk and two narrow
 stationary components identified with the inner edge of a
 circumbinary disk (Blundell, Bowler \& Schmidtobreick 2008). The striking change in the
 stationary H$\alpha$ complex occurred on or about JD +287.5, when the 
 H$\alpha$ complex broadened and clearly contains
 typically five components. Fig.2 of Schmidtobreick and Blundell (2006) also shows clear absorption notch structure on the blue side of the H$\alpha$ complex after JD +300 and in that same figure marked P Cygni characteristics are seen in He I spectra and very strongly in O I at 7772 \AA\ . It is worth noting that the moving H$\alpha$ line from the blue jet is to the blue of the stationary line on JD +291.5 and has transited to the red side on JD +294.5. On neither date did it confound stationary H$\alpha$.
 
 I therefore digitised the H$\alpha$ spectra plotted in Fig.2 of Schmidtobreick \& Blundell (2006), from JD +287 to 310 . These I fitted with five Gaussian profiles, occasionally introducing an absorption profile when it greatly improved the representation of the data. The fits  yield a
single broad emission line (standard deviation $\sim$ 20 \AA\  after JD +294) and  four
lines which are narrow (standard deviations of a few \AA\  only), Fig.1. In the period before the optical outburst the broad line was identified with the wind from the accretion disk and the lines which appear as the inner pair in the later period as coming from the circumbinary disk (Bowler, Blundell \& Schmidtobreick 2008). The results of my fitting show that the blue line from the circumbinary disk (approximately 6562 \AA\ ) can be traced as far as JD +291 but is reddened or indeed completely absorbed thereafter. The red circumbinary line is more robust and survives throughout the sequence, albeit with some reddening from time to time. This paper is concerned with the outer pair, which are clearly associated with the accretion disk because their source orbits the system with the compact object at a speed of approximately 175 km s$^{-1}$ and period of approximately 13 days.

 In Fig. 1 I show three examples of digitised spectra and the curves I fitted to them. The lowest panel is for JD +291, just after the compact object is receding at maximum speed. The circumbinary lines are clear, as is the red feature which complements the extreme blue. These extreme lines are probably present on JD +287 but become striking only on JD +288. The middle panel is for JD +294, when the compact object is eclipsed by the companion. In the upper panel, half a period later than the lowest, deviation blue has moved more than 6 \AA\  to approximately its bluest value. This shift is entirely consistent with a change of speed of approximately 350 km s$^{-1}$ (see Table~\ref{tab:one}), just as would be expected for the blue side of an accretion disk.

\begin{figure}[htbp]
\begin{center}
  \includegraphics[width=6cm]{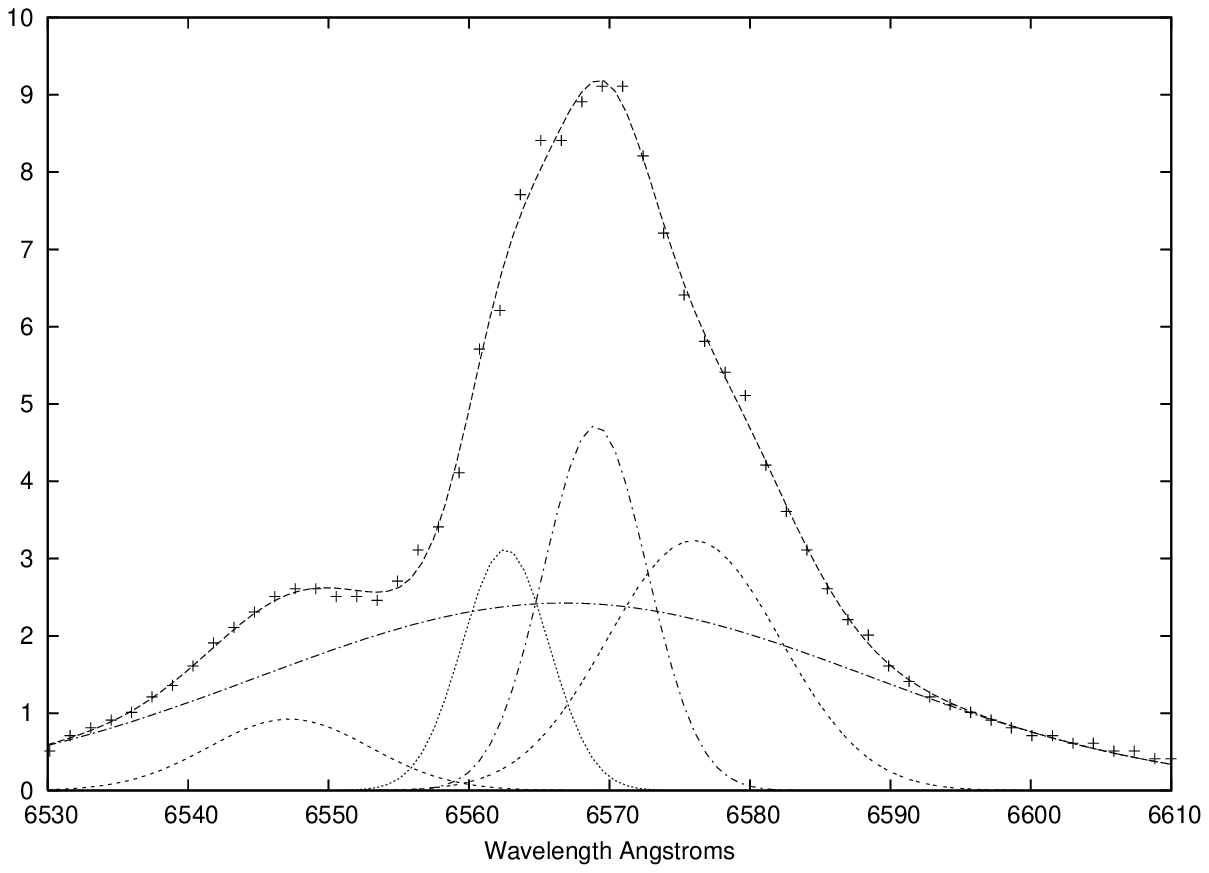} 
  \includegraphics[width=6cm]{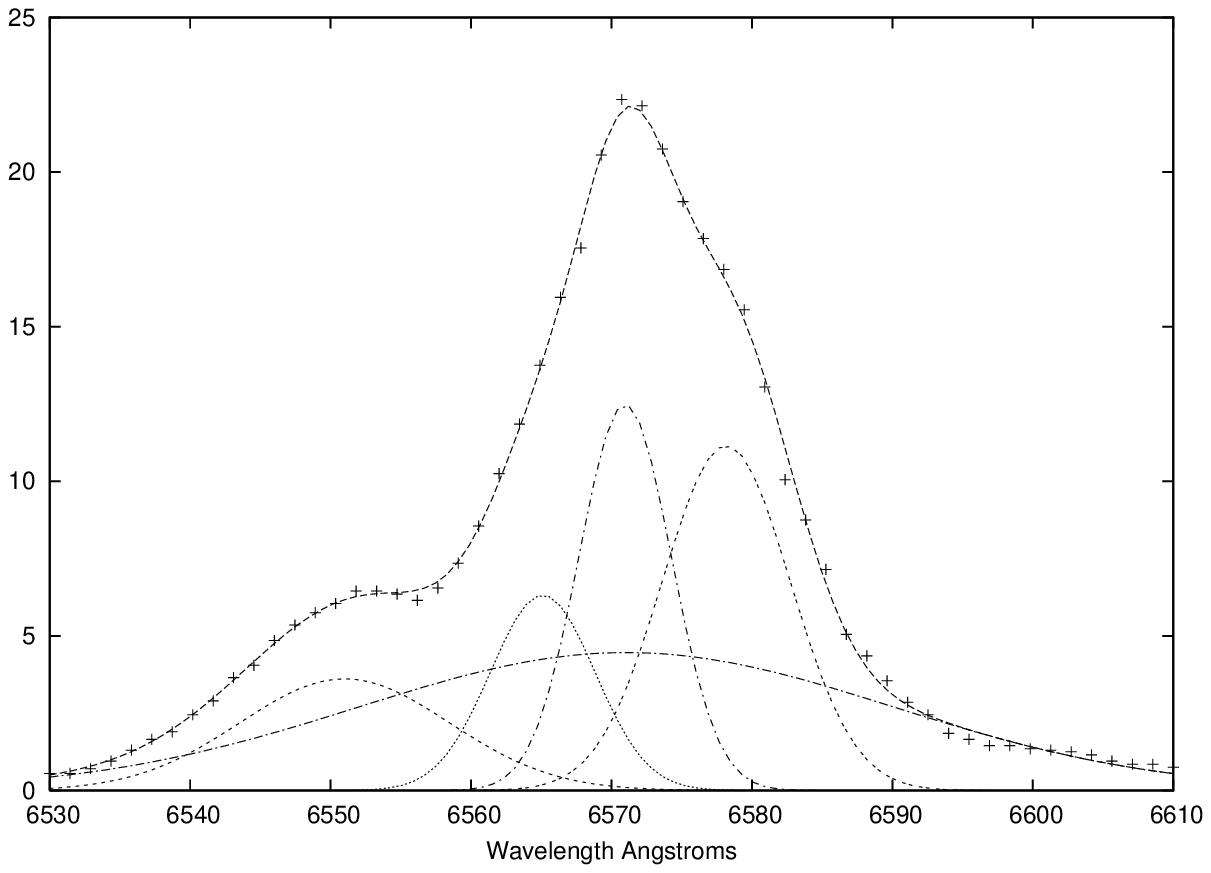}
 \includegraphics[width=6cm]{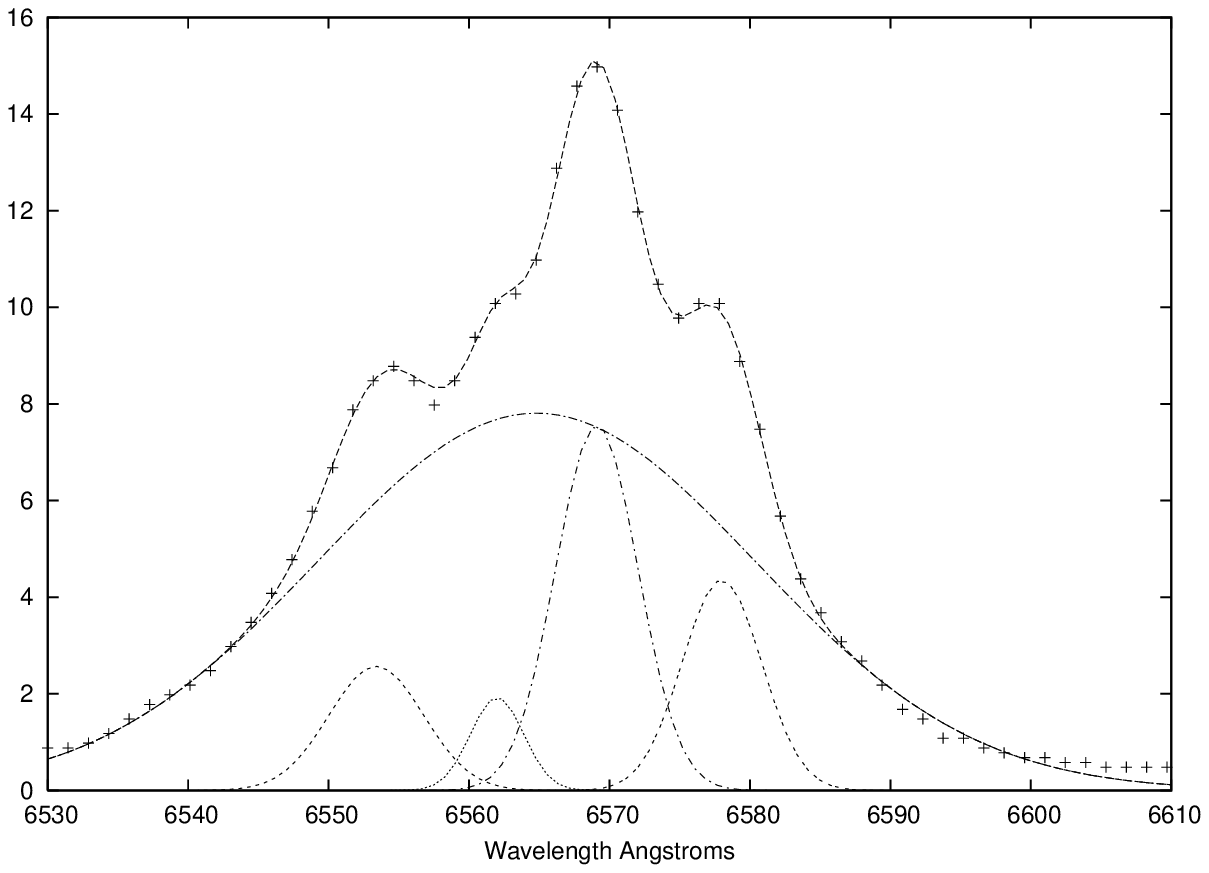}
 
   \caption{Three examples from the digitised
     sequence of H$\alpha$ spectra. The top panel is for JD +297.5, orbital phase 0.225, when the compact object is approaching. The middle panel is for JD +294.5, orbital phase 0.029 - an eclipse day. The bottom panel is for JD +291.5, orbital phase 0.796, when the compact object is receding. The fitted curves yielded for the wavelengths of the extreme blue component 6547.2, 6551.0 and 6553.4 \AA\  respectively. The extreme red feature which complements the blue is clear on JD +291 and becomes harder to track as it disappears into central region; see the large nominal errors for JD +297 in Table~\ref{tab:one}. On JD +291 the lines from the circumbinary disk are clear; absorption confounds the blue side at least as time progresses. }
\label{fig:ideagram}
\end{center}
\end{figure}

  The digitised spectra are the result of taking 56 samples from Fig.2 of Blundell \& Schmidtobreick (2006), with a spacing of 1.45 \AA. In most cases five Gaussian profiles were fitted; 15 parameters (see Table~\ref{tab:one}). The least squares fitting suffered from all the usual problems of errors frequently highly correlated and local minima in which a fitting program can easily get trapped. I explored the parameter space to the extent that I am confident that the central wavelengths of the exterior fitted Gaussians are usually determined to an accuracy of approximately 1- 2 \AA .  Examples of parabolic errors from the fitting program are also given in Table~\ref{tab:one}.

     It should be remembered that the relationship between the Doppler velocities determined from these centroids and the motion of bulk material in the environment of SS 433 may not be entirely straightforward. Very little is known about the accretion features of SS 433 and a phenomenological fit of this kind is the best that can be attempted at present. Strictly speaking, all that is certain is that there is a significant amount of radiating material moving so as to produce the projected Doppler velocity characterised by the centroid of any fitted Gaussian.

\begin{table}
\centering
\vspace{0.8cm}
\begin{tabular}{llrr}
\hline
 JD +291.5   &      h     & $ \lambda$ ( \AA\ ) & $\sigma$ (  \AA\ )   \\
 1     &2.57(0.82)   &6553.4(1.3)   & 3.4(0.4)       \\
 2     &1.93(1.87) &6562.0(2.1)  & 1.9(0.4)    \\
 3     &7.81(0.80)    &6564.8(0.3) &15.6(0.6)   \\
 4     &7.56(0.85)   &6569.1(0.3)   &2.9(0.2)  \\
 5     &4.36(0.70)   &6578.0(0.4)  &2.8(0.2)  \\
  \hline
 JD +294.5  &                            &               &                                        \\
 1     & 3.61(1.10)    &6551.0(1.5)  &7.5(0.9)   \\
 2     &6.32(5.59)    &6565.1(4.9)  &3.7(0.3)   \\
 3     &4.46(1.37)    &6571.0(2.2) &19.0(1.7)        \\
 4     &12.49(6.45)  &6571.0(2.0)  &3.3(0.2)  \\
 5     &11.15(4.54)  &6578.2(1.2)  &4.5(1.1)  \\
  \hline
 JD +297.5   &                     &                                        &                      \\
 1      &0.92(0.70) &6547.2(3.6)       &5.8(1.6)              \\
 2      &3.12(2.06)  &6562.6(2.6)    &3.1(2.6)     \\
 3      &2.43(0.96)  &6566.8(1.3)   &21.8(1.3)    \\
 4      &4.72(1.56)  &6569.0(1.9)   &3.7(1.9)   \\
 5      &3.23(1.96)  &6576.0(5.3)  &6.1(5.3)   \\
  \hline
  \end{tabular}
  \caption{\label{tab:one}  For each of the three spectra displayed in Fig.1 the fitted parameters of the five Gaussian components are listed. For each spectrum, the first column is the height h, in the units employed in Fig.1, the second column is the central wavelength $\lambda$ in \AA\, and the third is the standard deviation $\sigma$ of each Gaussian profile, also in \AA. The numbers in parentheses are the fitted errors.}
  \end{table}

\section{The accretion disk of SS433 unveiled}

Identification of the accretion disk of SS 433 depends on the complete set of data from JD +288.5 to + 310.5, almost two orbital periods. These data are displayed in different ways in Figs. 2 and 3. In Fig.2 the recessional velocities of the extreme red and blue features are plotted as a function of time in a format similar to that of Fig.1 of Blundell, Bowler \& Schmidtobreick (2008). Time increases upwards but is given in units of the orbital period, orbital phase 0 taken as eclipse of the compact object and its disk by the companion on JD + 281.

   Fig.2 shows that the blue feature is at its most red at JD +289.5, close to orbital phase 0.75, and has reached its extreme blue about seven days later. The difference in recession speeds of the blue source is about 350 km s$^{-1}$. The most plausible explanation is that the source of this feature is orbiting the centre of mass of the binary system with the compact object.  The fairly well
defined feature in the red exhibits  a similar pattern of deviation
toward the blue and after orbital phase 1.56 ( JD +301.5) both features are returning toward the red. The two lines, swinging together in phase, are separated by over 1000 km s$^{-1}$ and this is exactly the pattern to be expected from an accretion region orbiting the compact object, edge on, at over 500 km s$^{-1}$. During the period after JD +287 the jets were close to the plane of the sky, making an angle to the line of sight of $80^{\circ}$  initially, $85^{\circ}$on day 294 and closer to $90^{\circ}$ for the remainder of the observations; see Blundell, Bowler \& Schmidtobreick (2007).  If the H$\alpha$ emitting region of the accretion disk is perpendicular to the jet axis, then it is almost edge on during these observations and the Doppler speeds of the disk material, projected on the line of sight , are modulated by the orbital velocity of the compact object about the binary centre of mass. Further, it is clear that the two extreme features are drawing further apart with time, which strongly suggests that the material in the disk is speeding up. This increase in the difference of projected velocities is not likely to be due to the disk tilting; it is almost edge on during this period and the increase would require a tilt of $44^{\circ}$. Another possible effect is that different regions of the disk rim dominate as time goes on, as a result of drifting obscuring material.

For an ideal disk with material in a circular orbit about the compact object and no additional source, the  mean recessional speed of the red and blue components would trace a sinusoidal pattern following the orbit of the compact object about the binary centre of mass. The rotational speed of the material about the compact object would be given by half the difference of the recessional speeds. The latter is plotted in the upper part of Fig.3 and the former in the lower part; it is clear that the disk material is speeding up from about 500 km s$^{-1}$ to over 700 km s$^{-1}$ during almost two orbital periods. These numbers should probably be regarded as lower limits, applicable only if the regions of the disk rim to which the line of sight  is tangential dominate.

Figs. 2 and 3 also reveal that the simple picture of an ideal disk is unlikely to be wholly adequate. After orbital phase 1.25 (JD + 297) the blue line attributed to the edge of the disk does not return toward the red along a sinusoidal curve in Fig.2 but rather hangs in the blue for several days before snapping back rather abruptly. A similar pattern is observed for the red line after phase 1.75. These systematic departures from the most elementary model are readily explained if some radiation comes from a gas stream infalling to the rim of the accretion disk.

\begin{figure}[htbp]
\begin{center}
   \includegraphics[width=20cm]{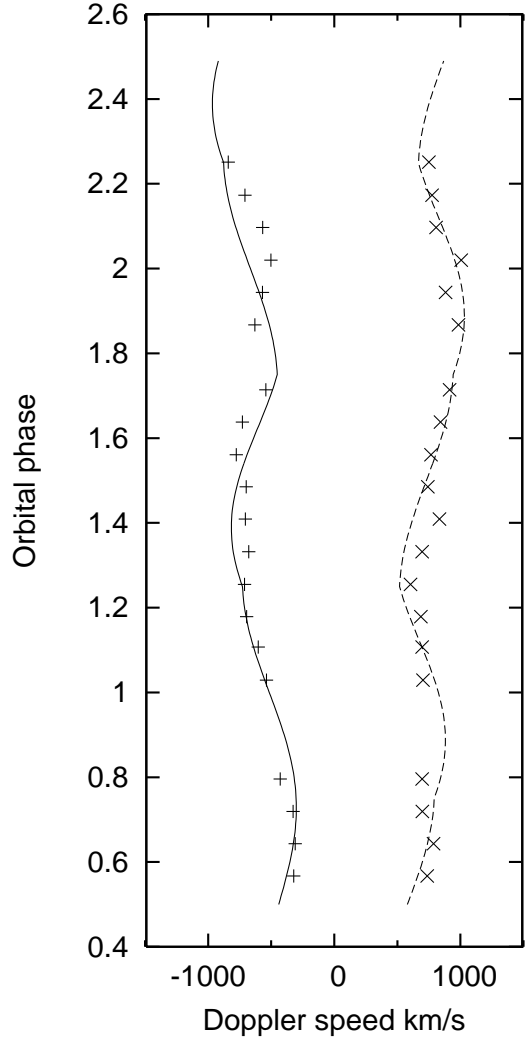} 
\caption{  The radial velocities of the extreme red and blue
components of H$\alpha$.  Although the data are barely adequate for the
purpose, they show an oscillation of amplitude consistent with 175 km s$^{-1}$, period 13 days, and a slowly increasing separation of the two components. The orbital phase is given in units of the period, with phase 1 corresponding to JD +294. The superimposed curves are for the simple model described in the text. }

\label{fig:disc}
\end{center}
\end{figure}

The curves superimposed on the data in Figs. 2 and 3 are for a simple model of a rotating disk which slowly speeds up, augmented by radiation from infalling gas travelling faster than the rim of the disk prior to joining it. The model is explained more quantitatively below, but it is clear that it provides a good explanation of the observations presented in Figs. 2 and 3. The data are somewhat erratic, but this may not be surprising in such a violent environment. The relation between the Gaussian centroids and the sources in the disk may not be straightforward and the disk is only intermittently visible in H$\alpha$ -  obscuring material  might have been moving around during the period of observation. One should beware of over parametrising the data, but the addition of an infalling stream accounts for the systematic deviations of the data from the simple model with only the disk, thereby adding verisimilitude to what might otherwise be held to be a bald and unconvincing narrative. Thus these data establish that the high speed components of the stationary H$\alpha$ lines which appear after JD +287.5 are contributed by a ring, or tight
spiral, within the accretion disk of SS 433. It is however not difficult to believe that the orbital motion of the binary might not be apparent in a substantially shorter sequence of observations, such as those in a recent study of the Brackett $\gamma$ line (Perez  M \& Blundell 2009).

\begin{figure}[htbp]
\begin{center}
 \includegraphics[width=9cm]{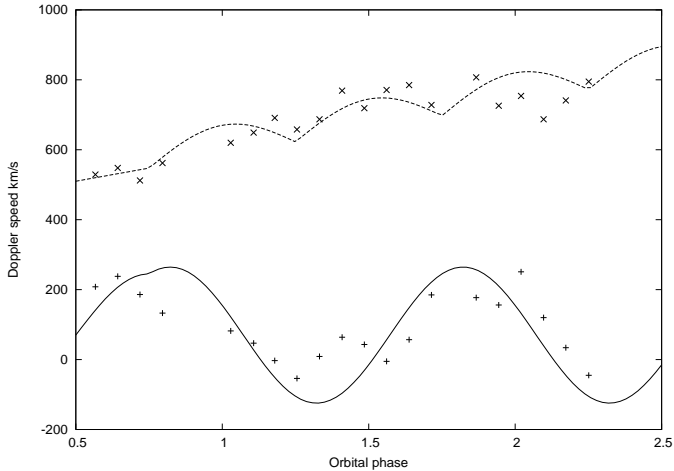}
 \caption{ The upper part of the figure shows half the difference between the recessional velocities of the red and blue components. This is approximately the rotational velocity of the accretion disk; that part dominating H$\alpha$ emission is increasing slowly over 20 days. The lower part of the figure shows the mean recessional velocity; approximately the velocity of the compact object about which the accretion disk is orbiting. The oscillation is approximately sinusoidal with a period of 13 days but the extremes are delayed a little beyond phase 0.25 (compact object approaching) and 0.75 (compact object receding). The superimposed curves are from the simple model discussed in the text.}
\label{fig:model}
\end{center}
\end{figure}

  The model discussed below assumes that an accretion stream becomes visible at or just after orbital phase 0.75.
 Figs. 2 and 3 show that the blue component is bluest just before a phase of 1.5 (rather than at 1.25) and similarly that red is reddest just before an orbital phase of 2. These are configurations where infalling gas on an orbit close to parabolic could present considerable depth along the line of sight and would be moving in the same sense as and faster than the orbiting material it is destined to join. I have modelled the effect of H$\alpha$ radiation from such a stream in Eqs.(1, 2) below. In the model the effective rotational speed of the disk rim {\it V} varies linearly from 510 km s$^{-1}$ at orbital phase 0.5 to 770 km s$^{-1}$ by JD +310; the orbital speed of the compact object about the centre of mass of the binary {\it $v_O$} is taken as 175 km s$^{-1}$ and an infalling stream has speed {\it$V +v_S$} in the frame of the compact object. The perceived speeds of the red and blue components are then given by

\begin{equation}
v_R = +V - v_O\sin\phi + S_R v_S \cos\phi
\end{equation}

\begin{equation}
v_B = -V - v_O\sin\phi + S_B v_S\cos\phi
\end{equation}

The switch function {\it$S_R$} is +1 if $\cos\phi$  $>$ 0 and zero otherwise; {\it$S_B$} is +1 if $\cos\phi$ $<$ 0 and zero otherwise. Thus for $\phi$ close to zero, when the companion eclipses the disk, the perceived velocity of the red source is {\it$V$} + {\it$v_S$} and of the blue rim  - {\it$V$}. Half a period later the gas stream is approaching and the perceived velocity of the red rim is + {\it$V$} and of the blue source -{\it$V$} - {\it$v_S$}.

   The dependence of the perceived speed of the gas stream on $\cos\phi$ and the switching functions is merely a simple representation of the physical argument;  the extremes are reached close to the times of eclipses and the effects vanish at elongations. The curves which are compared with the data in Figs. 2 and  3 were calculated from Eqs. (1, 2) with {\it$v_S$} set equal to 170 km s$^{-1}$ after orbital phase 0.75. A systemic red shift of 70 km s$^{-1}$ was added, taken from the apparent systemic red shift of the circumbinary disk (Blundell, Bowler \& Schmidtobreick 2008).
   
   The effective orbital speed of the disk material is about 500 km s$^{-1}$ when it is first revealed and slowly increases to 700 km s$^{-1}$ over almost two orbits. The widths of the accretion disk lines increase from standard deviation $\sim$ 3 \AA\  to 6 \AA\  over the same period. Around JD +287 either obscuring clouds were blown away or material was added to the outer edge of the accretion disk - or both; the increasing speed and width of the H$\alpha$ emission suggest a picture of matter added at the rim  spiraling deeper and faster into the interior. The disk was revealed in H$\alpha$ in violent times; this paper is not concerned with the flaring phenomenon except insofar as it affects understanding of the disk, but it is probably relevant that the speed of the wind increases rapidly from about 500 km s$^{-1}$ prior to JD +290 to around 1000 km s$^{-1}$ after JD +294.
   
    It is of some interest to establish a lower limit on the speed of a stable orbit: to this end compare the claim of Perez M \& Blundell (2009) to have observed the accretion disk in the far infrared Brackett $\gamma$ line. Those data were taken during a quiescent period; the effective orbital speed of the disk material is $\sim$ 500 km s$^{-1}$  and nowhere exceeds 600 km s$^{-1}$. [Although those data do not reveal the orbital motion about the centre of mass of the binary, it is hard to see what an alternative origin might be.] It is reasonable to take as a lower limit on speed of the outermost radiating hydrogen region  500 km s$^{-1}$. This lower limit represents the orbital speed if the accretion disk spectrum is dominated by regions to which the line of sight is tangential.

\section{The mass of the compact object and the size of the companion}

Knowledge of the speed of the compact object $v_{\rm X}$ about the centre of mass of the binary and the speed of the circumbinary disk determine a lower limit on $q$, the ratio of the mass of the compact object $M_{\rm X}$ to the mass of the stellar companion $M_{\rm C}$ [ a lower limit is $q$ $\sim$ 0.7, Blundell, Bowler and Schmidtobreick (2008)]. In the same way the observed speed of the accretion disk material sets, in principle, an upper limit on $q$. Because the system is binary, there is a critical radius $r_{\rm C}$ within which (Newtonian) orbits about the compact object are stable; for too great a mass $M_{\rm X}$ an orbit with speed 500 km s$^{-1}$ would lie too far out for stability. The quantity $r_{\rm C}$/$\it A$, $\it A$ being the separation of the two components of the binary, is an approximately linear function of $\it q$ only, varying from 0.2 for $\it q$=0.4 to 0.3 for $\it q$=1.4 (Paczynski 1977, Paczynski \& Rudak 1980, Szebehely 1980, Artymowicz \& Lubow 1994, Holman \& Wiegert 1999). The binary separation $\it A$ is given by $\it q$ and the radius of the orbit of the compact object about the centre of mass of the binary $r_{\rm X}$, in turn given by the period
and the  orbital speed $v_{\rm X}$ of the compact object. This important number is by now quite well known.  The He II result is $v_{\rm X}$ = 176 km s$^{-1}$ with an error of 13 km s$^{-1}$ ( see Fabrika 2004) and the C II result of Gies et al (2002) is 162 km s$^{-1}$ with an error of 29 km s$^{-1}$. Below I have taken the value of $v_{\rm X}$ to be 175 km s$^{-1}$  and $r_{\rm X}$ to be $3.15  \times 10^{12}$ cm.

   Disk material orbiting the compact object with speed $v_{\rm r}$, at radii $r$ less than $r_{\rm C}$ satisfies (approximately) the equation
   
  \begin{equation}
  v_{\rm r}^2 =\frac{GM_ {\rm X}}{r}
  \end{equation}
  
     It is convenient to substitute for $M_{\rm X}$ its value in terms of $v_{\rm X}$, the mass ratio $q$ and the binary separation $A$, thus obtaining
   
 \begin{equation}
 \left(\frac{v_{\rm r}}{v_{\rm X}}\right)^2 = \frac{q(1+q)}{r/A}
 \end{equation}
 
 If $r$ is set equal to the critical radius $r_{\rm C}$ the equation is solved for the maximum value of $q$ compatible with a given $v_{\rm r}$. With $v_{\rm r}$ identified with the orbital velocity at the critical radius and taken as 500 km s$^{-1}$,  Eq.(4) yields $q$ = 1.1, with the ratio $r_{\rm C}$ /$A$ = 0.28. The mass of the compact object $M_{\rm X}$ is then 37 $M_{\odot}$ and the mass of the system 71 $M_{\odot}$. These are of course upper limits for $v_{\rm r}$ 500 km s$^{-1}$- if the visible material is orbiting deeper than the limit of stability, as seems likely, then $r$ is less than $r_{\rm C}$ and given by Eq.(4) in terms of $v_{\rm r}$. The dependences of both $r$ and $r_{\rm C}$ on $\it q$ are listed in Table~\ref{tab:two}. This table covers the range of $q$ between the lower limit of Blundell, Bowler \& Schmidtobreick (2008) and the upper limit just derived. Since it appears that an incoming gas stream reaches in excess of 700 km s$^{-1}$ before plunging deep into the disk, the lower values of $q$ are favoured, because the material has to have far enough to fall to pick up speed.

 \begin{table}
 \centering
 \vspace{0.8cm}
 \begin{tabular}{lllllrrrr}
 \hline
 $q$       &$r/A$       &$r_{\rm C}/A$      &$a_{\rm r}/A$     &$a_{\rm b}/A$     &$r_{\rm L}/A$       &$M_{\rm X}$     &$M_{\rm S}$        &$A$      \\
  1.1       &0.14         &0.28                      &0.255                   &0.32                    &0.37                        &37                      &71
  &6.6       \\
  1.0       &0.125       &0.27                      &0.250                   &0.31                    &0.38                        &31                      &62
  &6.3           \\
  0.9       &0.105       &0.26                      &0.245                   &0.30                    &0.39                        &25                      &53
  &6.0           \\
  0.8       &0.09         &0.25                      &0.242                   &0.29                    &0.40                        &20                      &45
 & 5.7           \\
  0.7       &0.075       &0.24                      &0.240                   &0.28                    &0.41                        &16                      &38
  &5.4           \\
 \end{tabular}
 \caption{\label{tab:two} System parameters as a function of mass ratio $q$. The quantity $r$ is the radius at which the orbital disk  speed is 700 km s$^{-1}$; $M_{\rm X}$ the mass of the compact object. $a_{\rm r}$ is the upper limit to the radius of the companion for the red side of the disk to be visible on JD +307.5; $a_{\rm b}$ for the blue. $M_{\rm S}$ is the mass of the system, masses in units of $M_\odot$ . The binary separation $A$ is in units of 10$^7$ km.}
 \end{table}

   On JD +294 the companion eclipsed SS 433, closest conjunction occurring 0.38 days in advance of the observations, according to the Goranskii ephemeris. Both the red and blue components of H$\alpha$ were stable and neither were eclipsed ; both sides of the radiating disk apparently protruded beyond the limb of the companion. A second eclipse occurred on JD +307, this time 0.26 days in advance; again neither component was eclipsed. These observations set limits on the radius to which the companion is opaque, $a$. For any value of $q$, the speed of the orbiting material determines the radius of the orbit about the compact object and hence simple geometry sets a limit on the radius of the companion. If the companion were too big, it would obscure one or both of the projected edges of the accretion disk and it does not. Very roughly,  $a$/$A$ must be less than $\sim$ 0.3; for comparison, the Roche lobe radius is $\sim$ 0.4. The geometrical considerations are set out below (Eq.(5).
   
 The orbit is tilted at $12^{\circ}$ to the line of sight, the angle $i$ between the line of sight and the jet precession axis being $78^{\circ}$. The disk is close to edge on (precession phase 0.294 on JD +294, 0.374 on JD +307; the jets are very close to the plane of the sky) and the plane of the disk makes an angle  $\theta$ to the plane of the orbit, approximately $20^{\circ}$. The delay of 0.38 days corresponds to orbital phase 0.029 of a period ( $\phi$ = 0.181 rad) and 0.26 days to 0.02 of a period ($\phi$ = 0.126 rad). The eclipse of JD +307 is more restrictive than that of JD +294, because the disk extremes are orbiting at $\sim$ 700 km s$^{-1}$, at half the radius on JD +294. If $r$ is a disk radius, just touching the limb of the companion, then for the red side $r$/$A$ is related to $a$/$A$ by the condition

\begin{equation}
\left(\frac{r}{A}\right) \geq  x + \sqrt{ \left(\frac{a}{A}\right)^2 - y^2}
\end{equation}

where  $x = \sin\phi\cos\theta - \cos i \sin\theta$ and $y = \cos i \cos\theta + \sin\phi\sin\theta$. The equivalent relation for the blue side is obtained by reversing the sign of $x$. The numerical values of $x$ and $y$ for JD +307, 0.26 days after closest conjunction, are 0.047 and 0.238 respectively.  Table~\ref{tab:two} lists the orbital radii as a function $q$ for orbital speed 700 km s$^{-1}$ and the upper limits on the radius of the companion for the red side of the disk, $a_{\rm r}$, and for the blue side, $a_{\rm b}$, to be visible. Because the observations were made just after closest conjunction, the red side sets the stronger limits. The weaker constraint from the blue side is included because of the possibility that infalling material on the red side could be visible beyond the limb of the companion, even when the disk edge is eclipsed.

   It is frequently assumed that SS 433 is so luminous because the companion continually overflows its Roche lobe (Fabrika 2004). The observations of the accretion disc during eclipse cast some doubt on this assumption. The radius  $r_{\rm L}$ of the Roche lobe of the companion, in units of $A$, is a function only of the mass ratio $q$ (Eggleton 1983; see also Chanan, Middleditch \& Nelson 1976; note that these authors use the symbol $q$ to denote $M_{\rm C}$/$M_{\rm X}$). The values are given in Table~\ref{tab:two} and it is clear that admissible radii $a$ are smaller than the Roche lobe radii. Eq.(5) assumes a  simple disk geometry but it is likely that the volume of the companion opaque to H$\alpha$ does not fill its Roche lobe, at least not during this episode. This does not seem at odds with the relatively short optical eclipses (Antokhina \& Cherepashchuk 1987) but the values of $q$ extracted from longer X-ray eclipses, assuming that the companion fills its Roche lobe (Cherepashchuk et al 2005, 2008), are smaller than permitted by the data on the circumbinary and accretion disks. X-ray eclipses are very varied (Cherepashchuk et al 2008) and it may be that the X-rays are eclipsed not only by the limb of the companion but also by an atmosphere, associated gas streams and wind (Fabrika 2004, Cherepashchuk et al 2008). 
  
  These H$\alpha$ eclipse data suggest upper limits on the radius of the companion $a$/$A$. It is possible that the IR data of Perez M. \& Blundell (2009) set a lower limit. Those data show negligible intensity from the extreme blue component at an orbital phase of 0.96, half a day before maximum eclipse of the compact object by the companion. The precession phase is $\sim$ 0.5 and the geometry very simple; if this corresponds to the blue side of the accretion disk just disappearing behind the limb of the companion, then $a$/$A$ $\sim$ 0.21 - 0.23 for $q$ in the range 1.1 - 0.7. These lower limits compare well with the upper limits in Table~\ref{tab:one}.

   I note that Cherepashchuk et al (2005) have evidence for absorption lines associated with the companion and orbiting the binary centre of mass at 132 km s$^{-1}$. This corresponds to a value of $q$ of 0.75, hence a value for $M_{\rm X}$ of 18 $M_\odot$ and a companion mass of 24 $M_\odot$. These numbers are very close to those inferred from the properties of the circumbinary disk (Blundell, Bowler \& Schmidtobreick 2008) but were discarded by Cherepashchuk et al (2005) on the grounds that X-ray eclipse data are not consistent with a value of $q$ of 0.75 - if the companion fills its Roche lobe. 

\section{Conclusions}

  In this sequence of observations I have identified in H$\alpha$ the accretion disk of SS 433; it seems likely that Falomo et al (1987) had a glimpse in observations made in June 1979 and  both Dopita \& Cherepaschchuk (1981) and Kopylov et al (1985) in separate observations of the same episode made in July 1980.
    The centre of the accretion disk orbits the binary centre of mass at a speed consistent with 175 km s$^{-1}$, in agreement with the He II and C II measures of the orbital speed of the compact object about the centre of mass of the system.
    
      The speed about the compact object of the H$\alpha$ emitting material seems to increase with time from about 500 km s$^{-1}$ to perhaps as much as 700 km s$^{-1}$ near JD +310. The accretion disk was invisible up to day 287.5; either obscuring clouds blew away or material was added to the rim of the accretion disk and heated up. In a viscous disk, material speeds up as it spirals in to meet its fate. 
      
      The properties of the system now seem to be quite well known. The compact object is almost certainly a rather massive stellar black hole and the companion has mass between 20 and 30 $M_\odot$. The radius of the companion is probably $\sim$ 0.25 of the binary separation, $A$ $\sim$ $6 \times 10^7$ km. It is unlikely that the companion filled its Roche lobe during this episode of instability.

The sudden appearance of the accretion disk in H$\alpha$, the increase in wind speed some days later and the gradual increase in orbital velocity about
the compact object suggest that the instability was initiated by a
mass ejection from the companion, or some other major disturbance of the outer regions of the accretion disk.  

\begin{acknowledgements}
The results reported here are further evidence of the power of the Blundell/Schmidtobreick sequence of observations, made possible by the
grant of Director's Discretionary Time on the 3.6-m New Technology
Telescope. The spectra were originally fitted with Gaussians by K. M. Blundell and I was informed by her results and a number of stimulating discussions with her. I thank D. R. Bowler for patient instruction in the art of life after Fortran.
\end{acknowledgements}


\begin{thebibliography}{}

\bibitem[ ]{}
Antokhina, E. A. \& Cherepashchuk, A. M. 1987, Sov. Astron., 31, 295

\bibitem[Artymowicz \& Lubow(1994)]{ArtymoL1994}
Artymowicz, P., \& Lubow, S.~H.\ 1994, \apj, 421, 651


\bibitem[Blundell et al.(2007)]{BBS2007} Blundell, K.~M., 
Bowler, M.~G., \& Schmidtobreick, L.\ 2007, \aap, 474, 903 

\bibitem[Blundell et al.(2008)]{BBS2008} Blundell, K.~M., 
Bowler, M.~G., \& Schmidtobreick, L.\ 2008, \apjl, 678, L47

\bibitem[Chanan, Middleditch \& Nelson(1976)]{MiddleD76}
Chanan, G.~ A., Middleditch. J., \& Nelson, J.~E. \ 1976, \apj, 208,  512

\bibitem[Cherepashchuk et al.(2005)]{Chere2005}
Cherepashchuk, A.~M., et al. \ 2005, \aap, 437, 561

\bibitem[Cherespashchuk et al.(2008)]{Chere2008}
Cherepashchuk, A.~ M. et al. \ 2008,\  PoS 1 Nov  7th INTEGRAL workshop

\bibitem[Crampton et al.(1980)]{Cramptonetal1980} Crampton, D., Cowley, 
A.~P., \& Hutchings, J.~B.\ 1980, \apjl, 235, L131 

\bibitem[Crampton \& Hutchings1981]{CramptonH1981}
Crampton, D., \& Hutchings, J.~B.\ 1981, \apj, 251,  604

Dopita, M. A. ,  Cherepashchuk, A. M. 1981 Vistas in Astr. 25   51

Eggleton, P. P. 1983 Ap. J. 268  368

Fabrika, S.N., Bychkova,L.V.  1990 A\&A 240  L5

Fabrika, S.N., 1997 Astrophysics\&Space Sciences 252  439

Fabrika, S.  2004  Astrophysics\&Space Physics Reviews 12  1

\bibitem[Falomo et al.(1987)]{falamo1987} Falomo, R., Boksenberg, 
A., Tanzi, E.~G., Tarenghi, M., \& Treves, A.\ 1987, \mnras, 224, 323 

Gies, D. R. et al 2002 Ap. J. 566 1069

\bibitem[Goranskii et al.(1998)]{Goranskiietal98} Goranskii, V.~P., 
Esipov, V.~F., \& Cherepashchuk, A.~M.\ 1998, Astronomy Reports, 42, 209 

Holman, M. J. \& Wiegert, P. A., 1999, AJ, 117, 621

Kopylov, I. M. et al  1985  Sov. Astron. 29  186



Paczynski, B. 1977  Ap.J.  216, 822

Paczynski, B. \& Rudak, B. 1980 Acta Astron. 30, 237



Perez M., S. \& Blundell, K. M. 2009 MNRAS, 397, 849

Schmidtobreick, L. \& Blundell, K.\ 2006, VI Microquasar Workshop: 
Microquasars and Beyond

Szebehely, V. 1980, Celest. Mech., 22, 7





 \end{thebibliography}
\end{document}